\documentclass[a4paper]{article}

\usepackage{INTERSPEECH2020}

\usepackage{hyperref}
\usepackage{caption}
\usepackage{subcaption}

\title{Latent linguistic embedding for \\ cross-lingual text-to-speech and voice conversion}
\name{Hieu-Thi Luong$^{1,2}$, Junichi Yamagishi$^{1,2}$}
\address{
  $^1$National Institute of Informatics, Tokyo, Japan \\
  $^2$SOKENDAI (The Graduate University for Advanced Studies), Kanagawa, Japan
  }
\email{\{luonghieuthi,jyamagis\}@nii.ac.jp}
\begin{document}

\maketitle
\begin{abstract}
As the recently proposed voice cloning system, NAUTILUS, is capable of cloning unseen voices using untranscribed speech, we investigate the feasibility of using it to develop a unified cross-lingual TTS/VC system.
Cross-lingual speech generation is the scenario in which speech utterances are generated with the voices of target speakers in a language not spoken by them originally.
This type of system is not simply cloning the voice of the target speaker, but essentially creating a new voice that can be considered better than the original under a specific framing.
By using a well-trained English latent linguistic embedding to create a cross-lingual TTS and VC system for several German, Finnish, and Mandarin speakers included in the Voice Conversion Challenge 2020, we show that our method not only creates cross-lingual VC with high speaker similarity but also can be seamlessly used for cross-lingual TTS without having to perform any extra steps.
However, the subjective evaluations of perceived naturalness seemed to vary between target speakers, which is one aspect for future improvement. 

\end{abstract}

\section{Introduction}

Text-to-speech (TTS) and Voice Conversion (VC) are technologies that generate speech with a target voice from a text or a source utterance input.
Recent studies have shown that we can clone new voices for these speech generation systems using a small amount of speech data either with or without transcription \cite{luong2020nautilus} while maintaining a high quality and speaker similarity to the target speakers \cite{arik2018neural,liu2018wavenet}.
Voice cloning can be regarded as a classical scenario of speech generation in which the objective is to generate utterances that are as similar to the natural speech of the target speaker as possible.
This framing is straightforward and easy to evaluate, but it ignores the the potential of the speech synthesis system which is the ability to create something new and/or better than the original under certain terms and conditions.
Dysarthric speech reconstruction \cite{kain2007improving,wang2020end} and accent-reduction voice conversion \cite{aryal2014can,zhao2018accent} are two examples of scenarios in which we want to create a speech generation model with voices that are better than the originals while maintaining a certain level of speaker individuality.
Cross-lingual speech generation \cite{wu2008cross,sun2016personalized}, the scenario in which speech is generated in a language not spoken by the target speaker, is another one and it is also the main topic of this paper.

In the case of cross-lingual TTS, we could adapt an HMM-based TTS model to a target speaker whose data is not in the target language by establishing a phoneme mapping between the two languages \cite{wu2008cross,liang2010comparison,oura2010unsupervised}, training a bilingual state-sharing HMM model \cite{qian2009cross}, or factorizing the language and speaker components as transformation functions \cite{zen2012statistical}.
Similar principles have been applied to neural TTS systems. For example, several works have proposed training a multi-speaker multi-lingual neural TTS model capable of cross-lingual speech generation by utilizing factorizing speaker and language components \cite{li2016multi,fan2016speaker,zhang2019learning}.
Unsupervised speaker adaptation methods, such as speaker-adaptive TTS model conditioning on neural speaker embedding \cite{chen2019cross}, have also shown promising results for the cross-lingual scenario. Cross-lingual TTS is also the foundation for more interesting applications such as code-mixing speech synthesis \cite{rallabandi2019variational,cao2020code}.

In the case of cross-lingual VC, the systems are generally based on non-parallel VC systems, which are developed using the non-parallel utterances of source and target speakers.
Specifically, Phonetic PosteriorGram (PPG) based models are often used for the cross-lingual scenario \cite{sun2016personalized}.
Even though a monolingual PPG trained on the target language is good enough for cross-lingual VC, it was reported that a bilingual PPG \cite{zhou2019cross,cao2020code} or mixed-lingual PPG \cite{zhou2019modularized} can significantly improve the performance.
Non-parallel VC systems based on a Variational Autoencoder (VAE) \cite{mohammadi2018investigation} or Generative Adversarial Network (GAN) \cite{kameoka2018stargan,sisman2019study} are also applicable to the cross-lingual scenario.
Cross-lingual VC was also the main topic of the Voice Conversion Challenge 2020 (VCC2020) which is the basis of the evaluations we present in this paper.

In this study, we extend the versatile voice cloning method of the NAUTILUS system \cite{luong2020nautilus} to the cross-lingual speech generation scenario.
While we follow the VCC2020 theme, our primary focus is the development of a unified framework for both cross-lingual TTS and VC. The proposed system is expected to generate speech with target voices in a language that is not spoken by them by using either the TTS or VC input interface. Moreover, the performances when switching between the two modes should be relatively consistent.
In Section \ref{sec:methodology} of this paper, we describe our procedure to create a unified cross-lingual TTS and VC system for a target speaker. Section \ref{sec:experiments} provides details about the experimental setup, particularly the data conditions and in Section \ref{sec:evaluations} we discuss the results of the VCC2020 and those of our own listening test. We conclude in Section \ref{sec:conclusion} with a brief summary and mention of future works.

\begin{figure*}[t]

\begin{subfigure}[b]{0.31\textwidth}
         \centering
         \includegraphics[width=\textwidth]{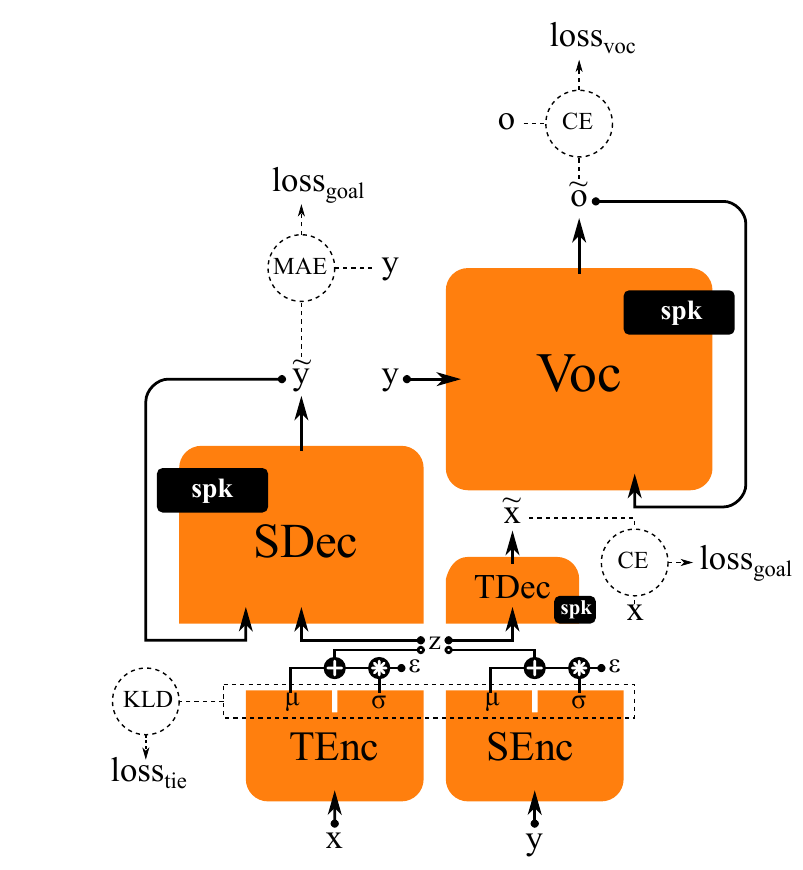}
         \caption{Initial training}
         \label{fig:step-training}
     \end{subfigure}
     \hfill
     \begin{subfigure}[b]{0.31\textwidth}
         \centering
         \includegraphics[width=\textwidth]{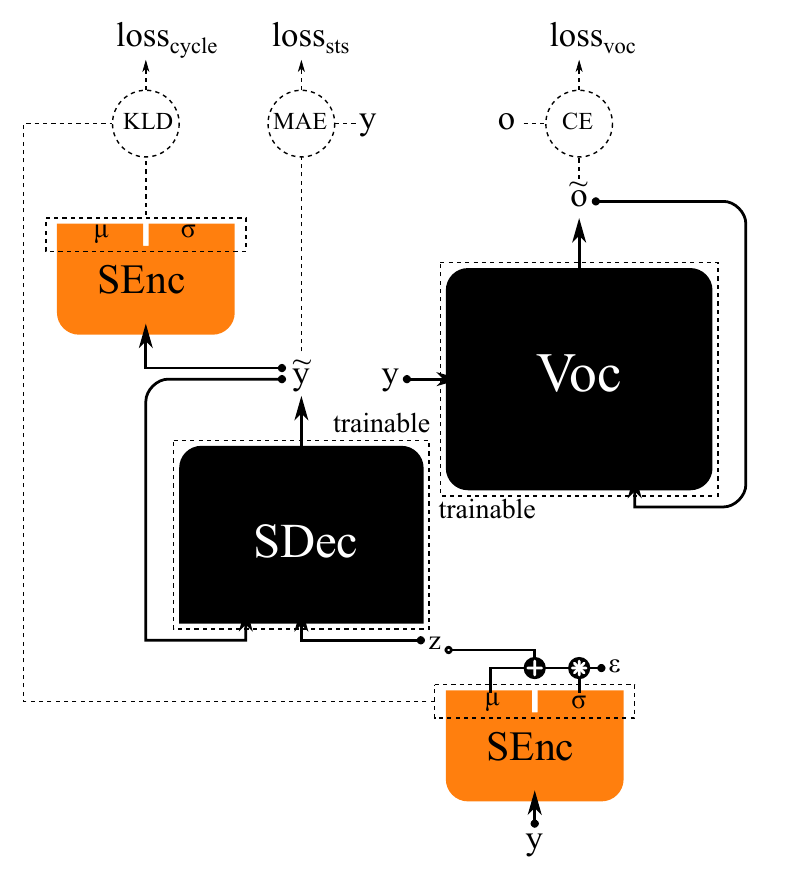}
         \caption{Step 1 - Adaptation}
         \label{fig:step-adaptation}
     \end{subfigure}
     \begin{subfigure}[b]{0.31\textwidth}
         \centering
         \includegraphics[width=\textwidth]{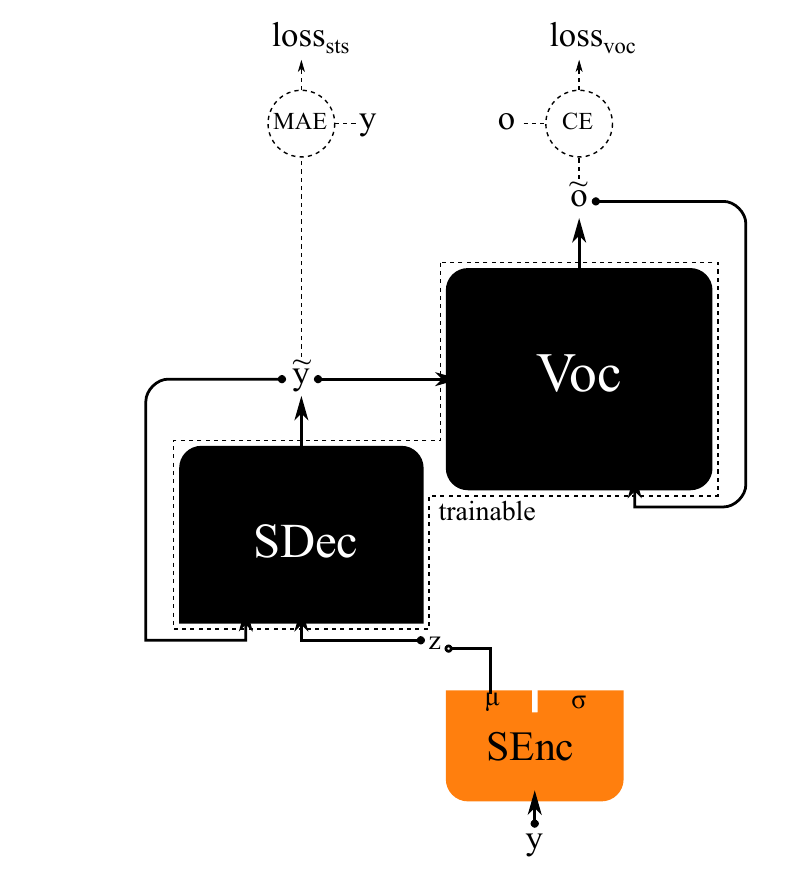}
         \caption{Step 2 - Welding}
         \label{fig:step-welding}
     \end{subfigure}
        \caption{Steps of the intra-language/cross-language speaker adaptation procedure for the proposed unified TTS/VC system. The proposed system comprises of a text encoder ($TEnc$), a speech encoder ($SEnc$), a text decoder ($TDec$), a speech decoder ($SDec$), and a neural vocoder ($Voc$). $\boldsymbol{x}$ is phoneme, $\boldsymbol{y}$ is acoustic, $\boldsymbol{o}$ is waveform, and $\boldsymbol{z}$ is latent linguistic embedding. The $\textrm{loss}_{goal}$ term is a placeholder for either $\textrm{loss}_{tts}$, $\textrm{loss}_{sts}$, $\textrm{loss}_{stt}$, or $\textrm{loss}_{ttt}$ depending on the encoder/decoder combination.}
        \label{fig:steps}
  \centering
\vspace{-3mm}
\end{figure*}

\section{Cross-lingual TTS and VC system with latent linguistic embedding}
\label{sec:methodology}

Our methodology for cross-lingual speaker adaptation is based on the unsupervised voice cloning strategy using Latent Linguistic Embedding (LLE) \cite{luong2020nautilus}.
Previously, we showed that this type of cross-lingual VC system can deal with several sub-scenarios thanks to the imbalance of data demands in each step \cite{luong2019bootstrapping}. In this work, we focus on the cross-language speaker adaptation scenario of TTS and VC to align with the theme of VCC2020.
The steps to develop a unified cross-lingual TTS/VC are summarized in this section, while further details can be found in the original paper \cite{luong2020nautilus}.

\subsection{Initial training}

This step focuses on training a robust and speaker-disentangled LLE in a target language as well as initializing all text/speech encoders/decoders using a large-scale multi-speaker corpus (Section III-A in \cite{luong2020nautilus}). We achieve this by jointly training the modules of the text-speech multimodal system using joint-goal and tied-layer objectives as follows:
\begin{equation}
\begin{aligned}
\label{eq:losstrain}
\textrm{loss}_{train} &= \textrm{loss}_{goals} +  \beta \; \textrm{loss}_{tie}  \\
                      &= \textrm{loss}_{tts} + \alpha_{sts} \; \textrm{loss}_{sts} + \alpha_{stt} \; \textrm{loss}_{stt} \\
                      &\qquad + \beta \; \textrm{loss}_{tie} \;.
\end{aligned}
\end{equation}
Given the VAE-like structure of the encoders, we use symmetrized Kullback-Leibler divergence between the encoder outputs as a tied-layer loss:
\begin{multline}
\label{eq:losstie}
    \textrm{loss}_{tie} = \frac{1}{2} \; L_{KLD}(TEnc(\boldsymbol{x}), SEnc(\boldsymbol{y})) \\ 
    + \frac{1}{2}\; L_{KLD}(SEnc(\boldsymbol{y}), TEnc(\boldsymbol{x})) \;.
\end{multline}
A multi-speaker WaveNet vocoder is separately trained using the same corpus in this step. It is used as the initial model for speaker adaptation in later steps.
\begin{equation}
    \textrm{loss}^\prime_{train} = \textrm{loss}_{voc} \; ,
\end{equation}
Figure \ref{fig:step-training} shows the conceptual actions performed in the initial training step. The same initial model obtained in this step will be reused for all target speakers.

\subsection{Step 1 - Adaptation}

As cross-lingual adaptation can be considered a special case of unsupervised speaker adaptation in which the transcribed speech of target speakers is unobtainable, as it is in a foreign/unseen language, we can use the unsupervised voice cloning strategy to tackle the cross-lingual scenario (Section III-B1 in \cite{luong2020nautilus}).
Specifically, we adapt the speech decoder to the speech data of the target speaker as illustrated in Fig.\ \ref{fig:step-adaptation}:
\begin{equation}
    \textrm{loss}_{adapt} = \textrm{loss}_{sts} + \beta \; \textrm{loss}_{cycle} \; ,
\end{equation}
where $\textrm{loss}_{cycle}$ is the KL divergence between the LLE distributions extracted from natural and converted acoustic features. As LLE is a latent representation of sound, it is expected to be generalizable to other languages. The neural vocoder is also adapted to the target speaker, as
\begin{equation}
    \textrm{loss}^\prime_{adapt} = \textrm{loss}_{voc} \; .
\end{equation}
The fact that the initial model was never trained on data in the languages other than English might negatively affect the performance of the cross-language speaker adaptation.
However, we decided not to use additional data of the languages spoken by the target speakers in order to keep data demand low and simply test the generalization ability of the LLE.

\subsection{Step 2 - Welding}

Although tuning the speech decoder and neural vocoder separately is sufficient, we perform an additional step in which they are jointly tuned in order to increase their compatibility, as showned in Fig.\ \ref{fig:step-welding}. The optimizing loss is formulated as follows:
\begin{equation}
    \textrm{loss}_{weld} = \textrm{loss}_{sts} + \gamma \; \textrm{loss}_{voc} \; .
\end{equation}
The $\textrm{loss}_{sts}$ is included to maintain the acoustic space for the autoregressive speech decoder. Several training tactics are applied in this step to enable the model to learn fine-grained details while avoiding overfitting \cite{luong2020nautilus}. 

\subsection{Step 3 - Inference}

The adapted speech decoder and neural vocoder can be used along with either the text or speech encoder to form a complete TTS or VC system. 
For the standard intra-language speaker adaptation scenario, the NAUTILUS system has demonstrated a highly consistent performance between the TTS and VC inference \cite{luong2020nautilus}. In this work, we check if the same statement holds true in the cross-lingual scenario.

\begin{figure*}[t]
\begin{subfigure}[b]{0.45\textwidth}
         \centering
         \includegraphics[width=\textwidth]{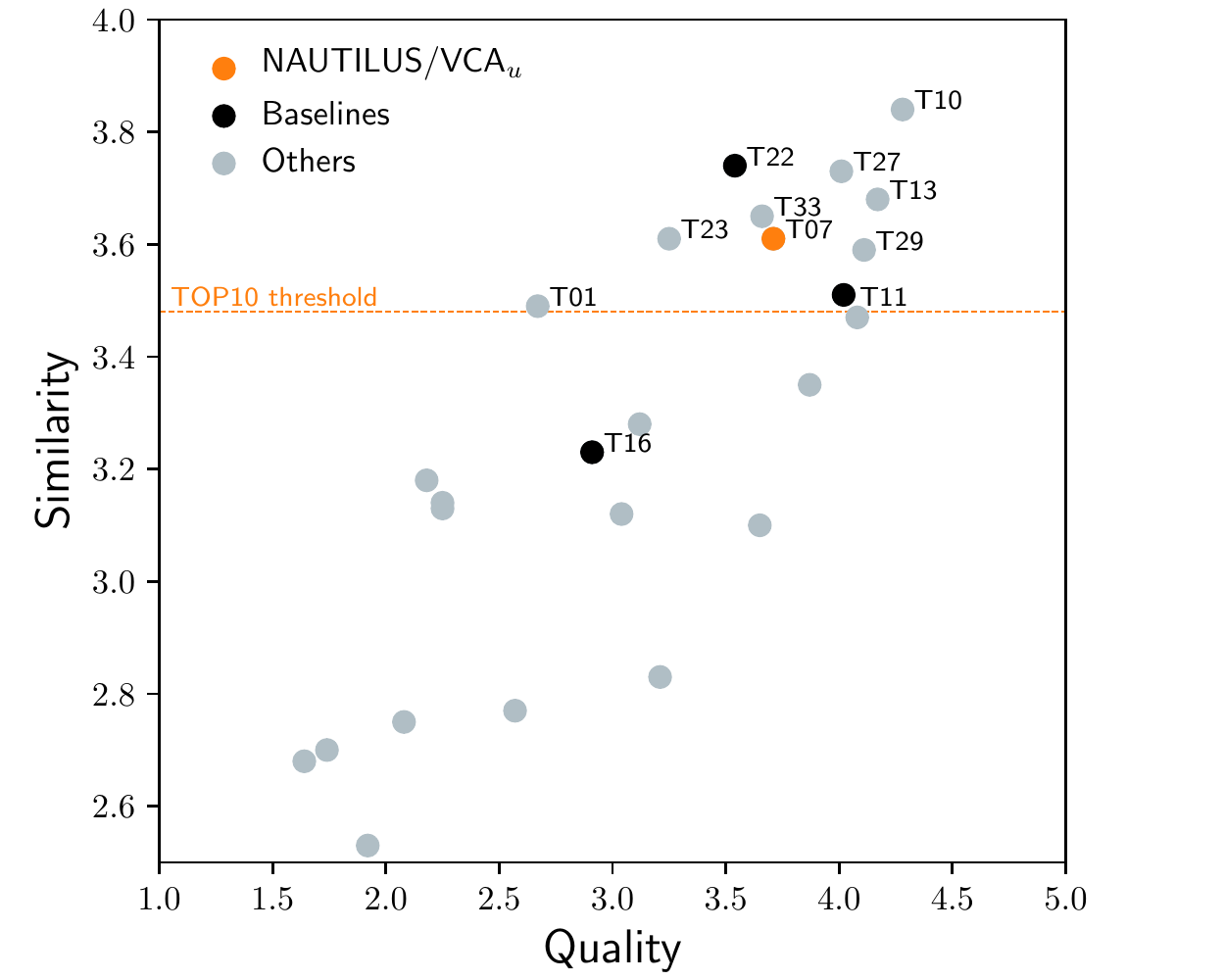}
         \caption{Standard scenario (Task 1)}
         \label{fig:result-task1}
     \end{subfigure}
     \hfill
     \begin{subfigure}[b]{0.45\textwidth}
         \centering
         \includegraphics[width=\textwidth]{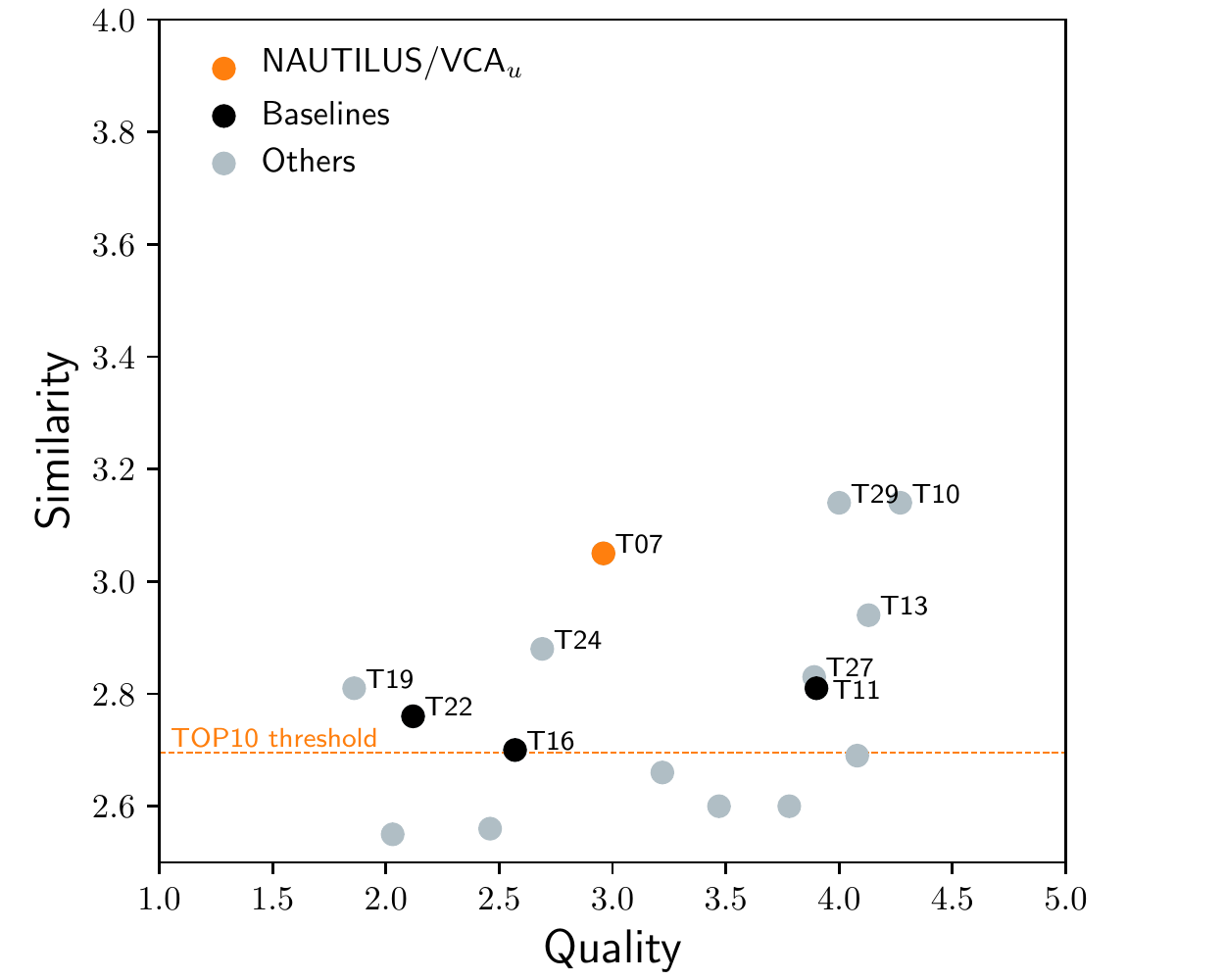}
         \caption{Cross-lingual scenario (Task 2)}
         \label{fig:result-task2}
     \end{subfigure}
        \caption{Subjective evaluations of quality and speaker similarity at VCC2020. Listeners are native English speakers.}
  \centering
\vspace{-3mm}
\end{figure*}

\section{Experiments}
\label{sec:experiments}

\subsection{Model configuration and LLE training}

The same NAUTILUS system used to evaluate scenario B in the original paper \cite{luong2020nautilus} was reused for the cross-language speaker adaptation experiments.
Each module of the neural model consisted of several one-dimensional convolution layers to capture temporal context. Phonemes were used as the text representation and an 80-dimensional mel-spectrograms were used as the acoustic representation.
The current system is not yet an end-to-end (E2E) system, as it requires the explicit phoneme duration information when generating speech from text input. This setup allows us to have the same duration condition between the TTS and VC system.
The initial model used for scenario B (and reused in this paper) was first trained on 24 kHz English transcribed speech of LibriTTS \cite{zen2019libritts}, which has diverse linguistic content, and then on VCTK corpus \cite{veaux2017superseded}, which recorded in a more controlled environment. This initial model trained on English data was adapted to target speakers whose data is in languages other than English.
Additional data of languages spoken by the target speakers would likely improve the performance \cite{zhang2019learning,zhou2019cross}, but we leave such investigation for future works.

\subsection{Target speakers for evaluations}

We tested the performance of the proposed cross-lingual speech generation system on target speakers from VCC2020, which consists of four native English speakers (Task 1) and six speakers who spoke languages other than English (Task 2).
While the focus of the work is the cross-lingual scenario, we perform both tasks to establish a reliable baseline with English target speakers. In summary, the target speakers included four English, two Finnish, two German, and two Mandarin speakers with an equal number of male and female speakers in each language. Each speaker had 70 utterances (around five minutes) of untranscribed speech available for adaptation.
We tested our cross-lingual TTS system on the same target speakers and compared its performance with our cross-lingual VC system to evaluate the consistency between the two.
Even though the challenge provided speech data of source speakers, we decided not to use it for either training or adaptation.

\section{Evaluations}
\label{sec:evaluations}

\subsection{Voice Conversion Challenge 2020}

Figure \ref{fig:result-task1} and \ref{fig:result-task2} shows the subjective quality and speaker similarity results, judged by native English speakers, of the systems submitted to VCC2020. We recreated the figures based on results provided by the organizers to focus on the information relevant to our works\footnote{See the paper written by the organizers for detailed results \cite{zhao2020voice}}.
For the standard intra-language task, our system, T07, achieved moderate performance and ranked among the top ten systems with the highest speaker similarity (Fig.\ \ref{fig:result-task1}).
Our system had a higher speaker similarity but lower quality than T11, which was the best system (N10) of VCC2018 \cite{liu2018wavenet}. This is similar to the results presented in the original paper of the NAUTILUS system \cite{luong2020nautilus}.
Compared with other baselines, our system had a higher quality but lower speaker similarity than T22 \cite{Huang2020}, while T16 \cite{tobing2019non} did not make it into the top ten ranking in the intra-language scenario. In summary, the subjective results of Task 1 confirm the competitive performance of our system when used as a VC system.

\begin{figure*}[t]
\begin{subfigure}[b]{0.33\textwidth}
         \centering
         \includegraphics[width=\textwidth]{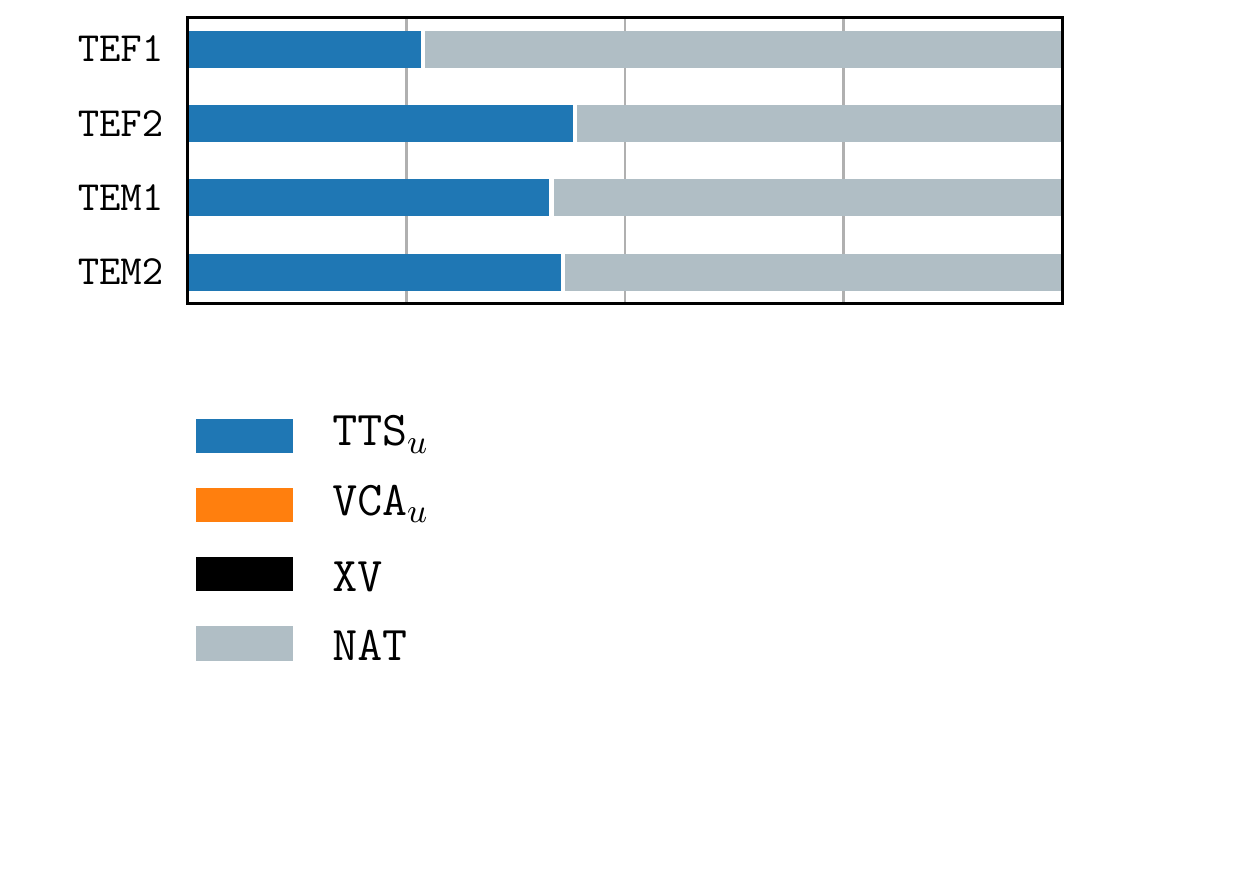}
         \caption{TTS$_u$ vs. NAT}
         \label{fig:quality-ttsnat}
     \end{subfigure}
     \hfill
     \begin{subfigure}[b]{0.33\textwidth}
         \centering
         \includegraphics[width=\textwidth]{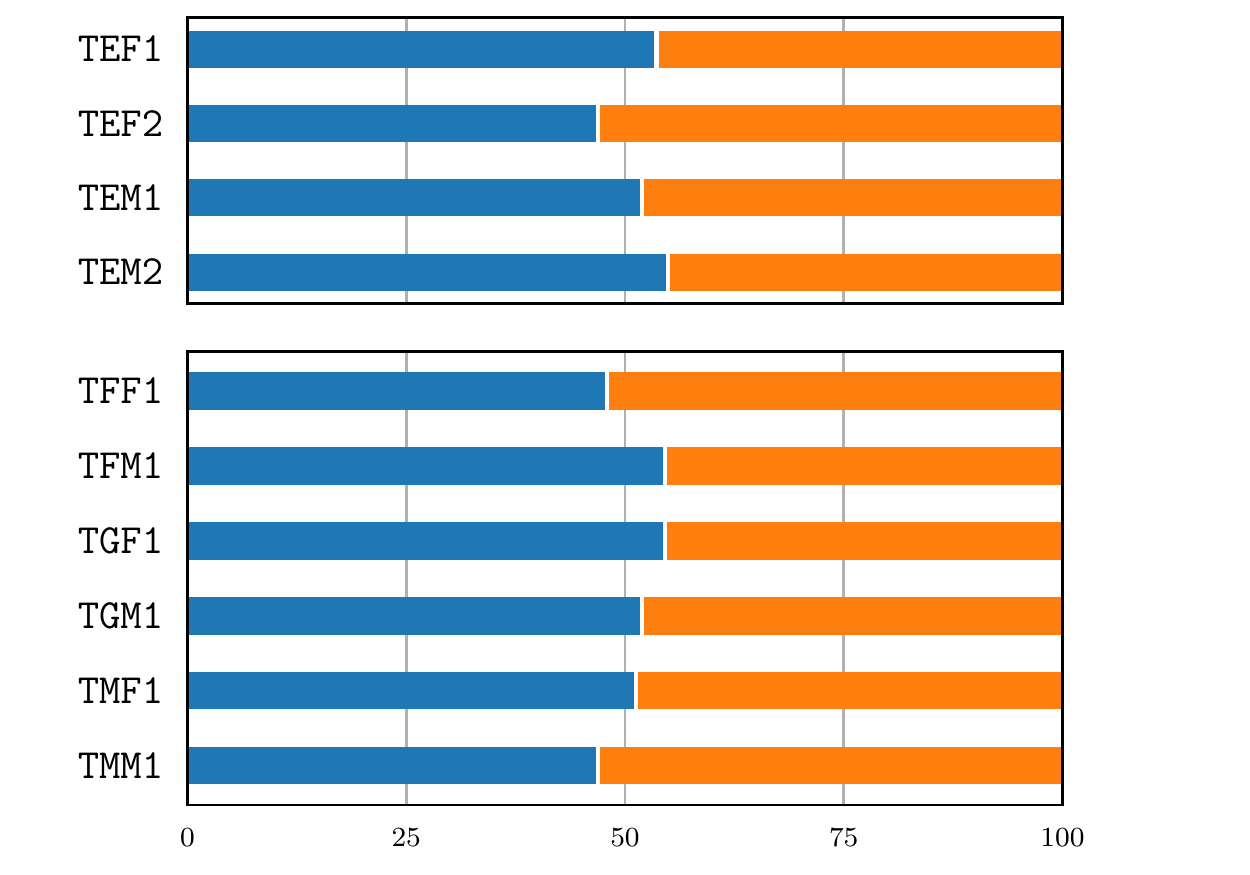}
         \caption{TTS$_u$ vs. VCA$_u$}
         \label{fig:quality-ttsvca}
     \end{subfigure}
     \hfill
     \begin{subfigure}[b]{0.33\textwidth}
         \centering
         \includegraphics[width=\textwidth]{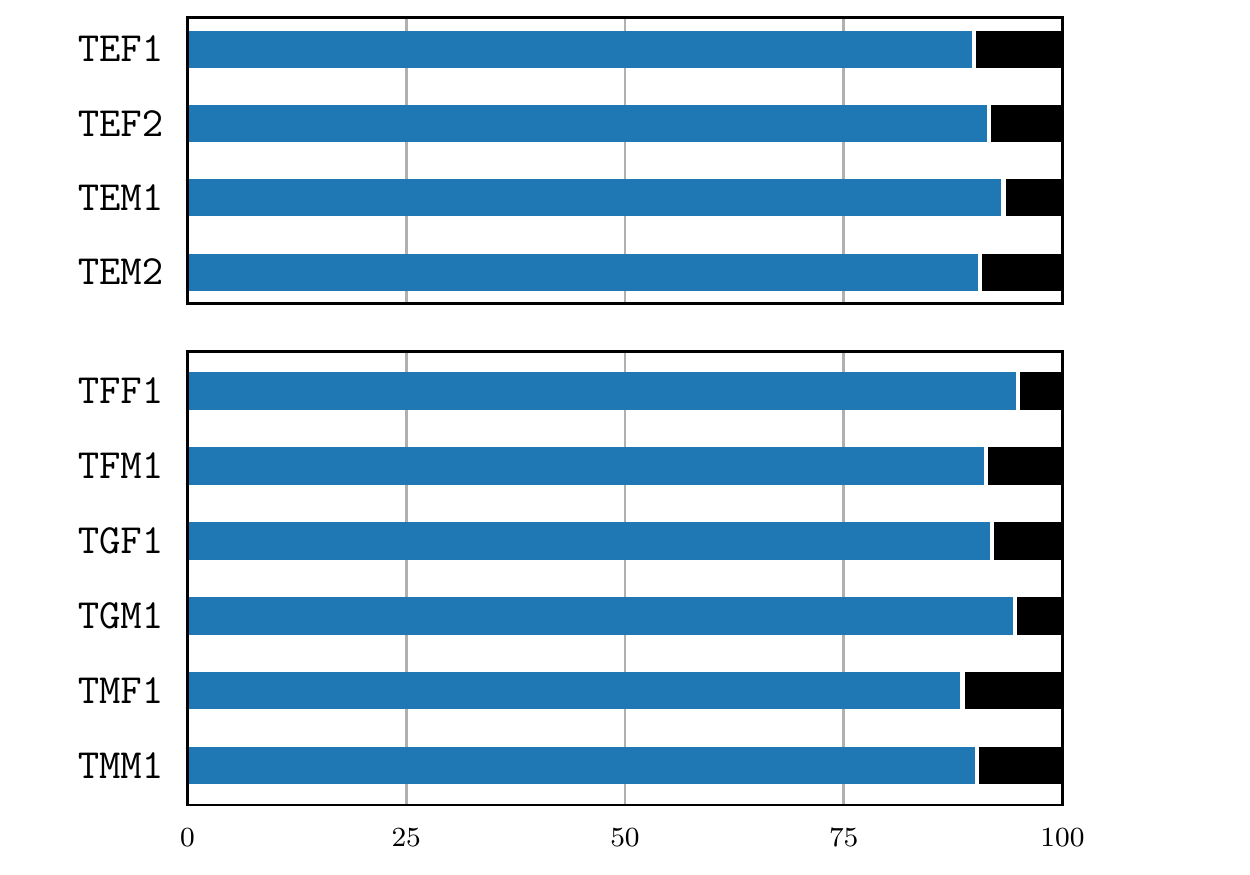}
         \caption{TTS$_u$ vs. XV}
         \label{fig:quality-ttsxv}
     \end{subfigure}
     \vspace{-5mm}
        \caption{Subjective quality evaluations. Each speaker/system pair was judged 300 times by native Japanese speakers.}
        \label{fig:quality}
  \centering
\end{figure*}

\begin{figure*}[t]
\begin{subfigure}[b]{0.33\textwidth}
         \centering
         \includegraphics[width=\textwidth]{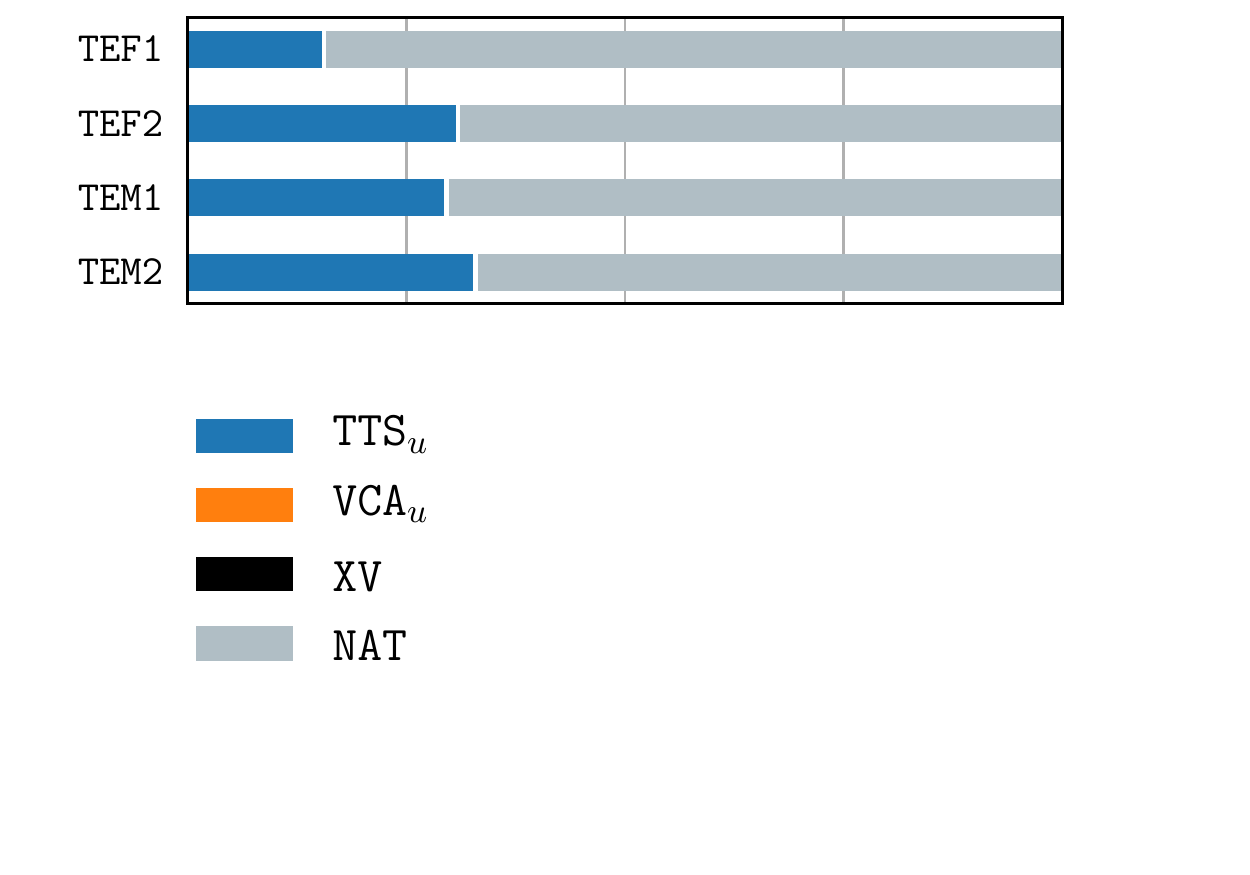}
         \caption{TTS$_u$ vs. NAT}
         \label{fig:similarity-ttsnat}
     \end{subfigure}
     \hfill
     \begin{subfigure}[b]{0.33\textwidth}
         \centering
         \includegraphics[width=\textwidth]{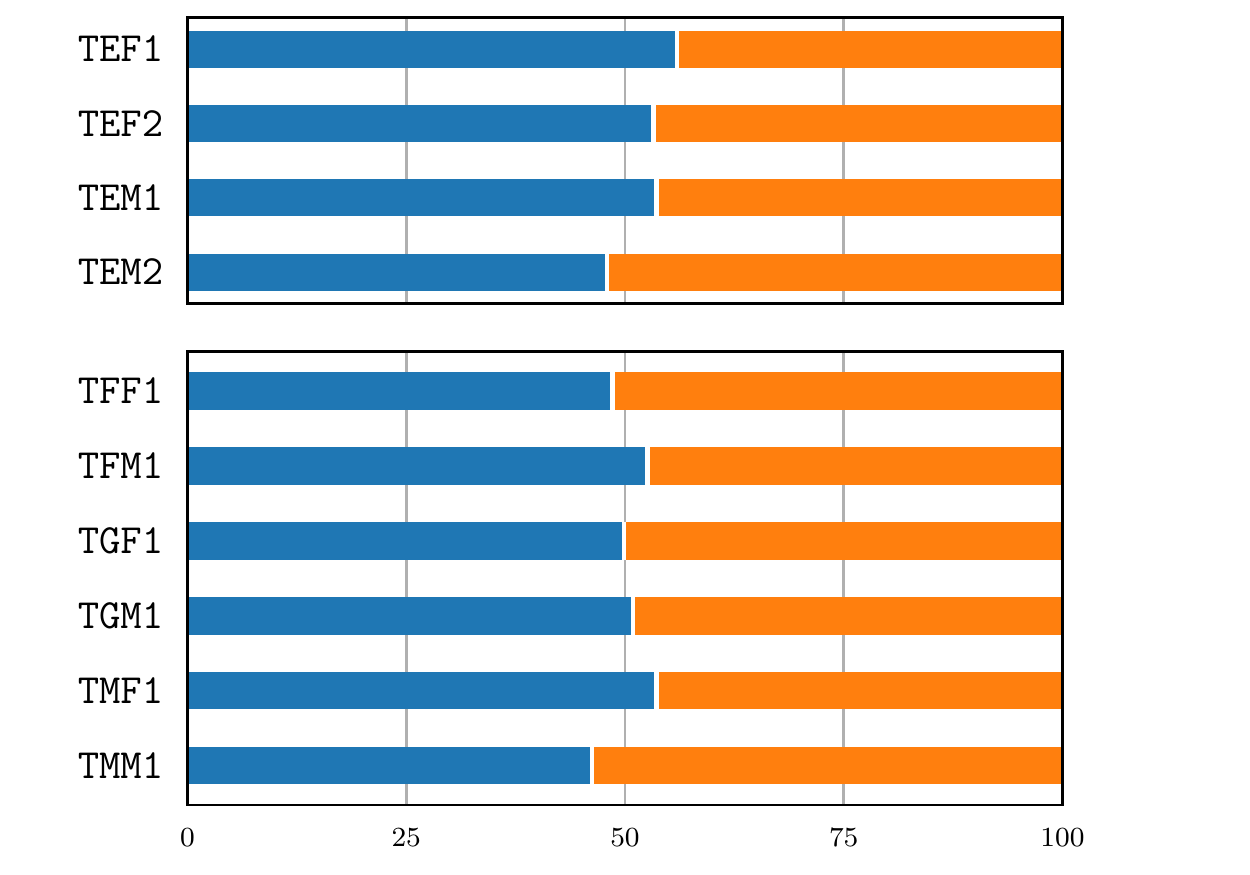}
         \caption{TTS$_u$ vs. VCA$_u$}
         \label{fig:similarity-ttsvca}
     \end{subfigure}
     \hfill
     \begin{subfigure}[b]{0.33\textwidth}
         \centering
         \includegraphics[width=\textwidth]{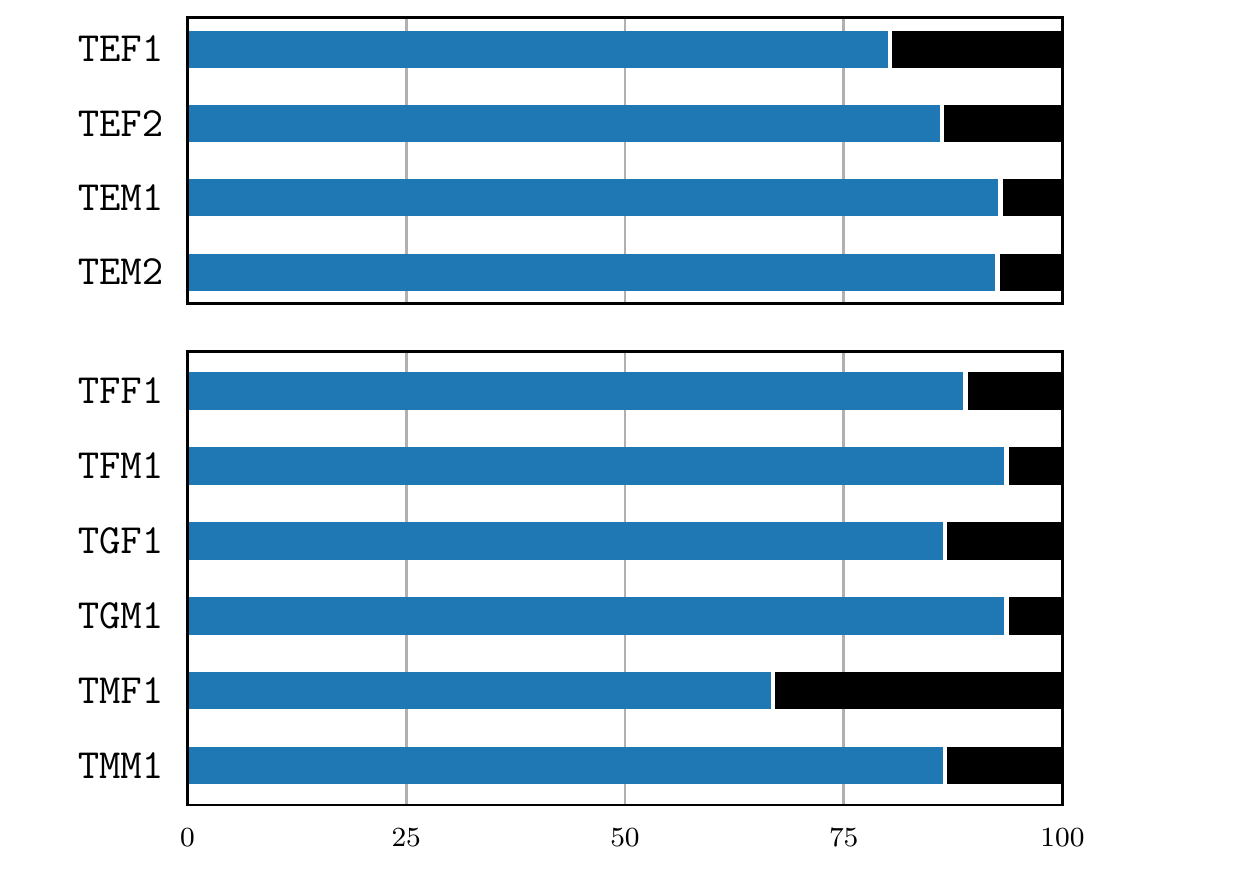}
         \caption{TTS$_u$ vs. XV}
         \label{fig:similarity-ttsxv}
     \end{subfigure}
          \vspace{-5mm}
        \caption{Subjective similarity evaluations. Each speaker/system pair was judged 300 times by native Japanese speakers.}
        \label{fig:similarity}
  \centering
\vspace{-3mm}
\end{figure*}

For the cross-lingual task, the setup for the listening test was trickier due to the language mismatch betweeen the generated and natural utterances. 
Based on information provided by the organizers, for the speaker similarity evaluation, listeners were presented with either a native language utterance or the English utterance of the target, the data of which was not provided to the challenge participants.
The results shown in Fig.\ \ref{fig:result-task2} are the averages over both cases.
In general, the cross-lingual scenario seems to have had a lower speaker similarity than the standard scenario, which is not surprising.
Our system (T07) was ranked 3rd in the similarity metric and higher than all three baselines.
However, it received relative low quality scores compared to other systems.
Interestingly, if we only consider the speaker similarity results between converted utterances and the native language utterance of the target speakers, our system ranked 1st in the case of Finnish speakers, 4th in the case of German speakers, and 4th in the case of Mandarin speakers. 
This result demonstrates the feasibility of using the NAUTILUS system for cross-language speaker adaptation, as it achieved very high speaker similarity without having to pre-train on data of languages spoken by the target speakers.

\subsection{A unified cross-lingual system of TTS and VC}

The strength of the NAUTILUS system is its ability to switch between TTS and VC modes. We evaluate its performance as a cross-lingual TTS system in this section.
Specifically, we compared our cross-lingual TTS system (TTS$_u$) with the cross-lingual VC system (VCA$_u$), a simple TTS baseline based on x-vector (XV) \cite{chen2019cross}, and the natural utterances of the target speakers (NAT)\footnote{Samples are available at \url{https://nii-yamagishilab.github.io/sample-cross-lingual-tts-vc/}}.
The adapted models used in the previous section were reused here for TTS$_u$ and VCA$_u$, while XV\footnote{XV system consists of the pretrained \textit{libritts.tacotron2.v1} and \textit{libritts.wavenet.mol.v1} models based on ESPnet toolkit \cite{watanabe2018espnet}} was the same system used in the original paper \cite{luong2020nautilus}.
We conducted our own listening test by asking 170 participants, who did one to ten tests, to pick which of the two samples had better quality or similarity. 
Each test consisted of 24 quality and 24 similarity questions covering all speaker and system pairs. For the TTS$_u$/NAT pair, we only evaluated the English target speakers, as comparing an English and non-English utterances would be uninformative.
The reference utterances used in the similarity question were in the native languages of the target speakers.

The quality and speaker similarity results are presented in Figures \ref{fig:quality} and \ref{fig:similarity}, respectively. As expected TTS$_u$ was not comparable with the natural utterances, but neither was it totally dominated by NAT, which is an encouraging result. Interestingly, the listeners seemed to pick the natural samples more when presented with a reference utterance as shown in the similarity results.
Between TTS$_u$ and VCA$_u$, there were no clear winners across all speakers which suggests the highly consistent performance between the two modes of the proposed system even in the cross-lingual scenario.
Last but not least, compared with the cross-lingual TTS baseline, XV, our system dominated the results across all speakers, which validates our cross-lingual methodology on the TTS side. If we look at individual speakers, our system seems to have had noticeably worse results in the case of TEF1 and TMF1 in the similarity metric. This suggests that the performance of the proposed method may not be consistent between speakers.

\section{Conclusions}
\label{sec:conclusion}

The cross-lingual experiments presented in this paper have once again demonstrated the versatility of the proposed NAUTILUS system.
Even though the initial LLE is only trained on English speech data, the proposed system can generalize to other languages spoken by the target speakers and generate utterances with high speaker similarity. This significantly reduces the data demands on the cross-lingual system and allows it to adapt to target speakers who only speak low-resource languages.
However, as cross-lingual systems are expected to benefit from multilingual data \cite{chen2019cross,zhou2019cross}, our future work will focus on taking advantage of either transcribed or untranscribed speech from large-scale multilingual corpora.
Moreover, although the VCC2020 has taken the first step in shifting the focus of the research community to developing speech generation system that are better and different from the original voice, concrete guidelines on how to evaluate such systems have not yet been established. Additional research on this new and exciting topic is required.

\section{Acknowledgements}

This work was partially supported by a JST CREST Grant (JPMJCR18A6, VoicePersonae project), Japan, and MEXT KAKENHI Grants (16H06302, 17H04687, 18H04120, 18H04112, 18KT0051), Japan.

\bibliographystyle{IEEEtran}

\bibliography{main}

\end{document}